\documentclass[article,superscriptaddress,onecolumn,preprint]{revtex4-1}

\usepackage{graphicx}% Include figure files
\usepackage{dcolumn}% Align table columns on decimal point
\usepackage{bm}% bold math

\makeatletter
\renewcommand\@make@capt@title[2]{%
  \@ifx@empty\float@link{\@firstofone}{\expandafter\href\expandafter{\float@link}}%
   {\textbf{#1}}\@caption@fignum@sep#2\quad
}%
\makeatother

\usepackage{amssymb}
\usepackage{amsfonts}
\usepackage{amsmath}
\usepackage{textcomp}
\usepackage{epstopdf}
\usepackage{wasysym}
\usepackage{xcolor}
\usepackage{braket}
\usepackage[colorlinks]{hyperref}
\hypersetup{
                pdfstartview={FitH},
                linkcolor=blue,
                citecolor=blue,
                filecolor=blue,
                urlcolor=blue
}

\begin{document}

%\preprint{APS/123-QED}

%\title{SCI by quantum confinement\\ in ultrathin single crystalline bismuth grown on Ge(111)}% Force line breaks with \\

\title{Tuning spin-charge interconversion\\ with quantum confinement in ultrathin Bi/Ge(111) films}% Force line breaks with \\

\author{C.~Zucchetti}
\affiliation{LNESS-Dipartimento di Fisica, Politecnico di Milano, Piazza Leonardo da Vinci 32, 20133 Milano, Italy}

\author{M.-T~Dau}
\affiliation{Univ. Grenoble Alpes, CEA, CNRS, Grenoble INP (Institute of Engineering Univ. Grenoble Alpes), INAC-Spintec, 38000 Grenoble, France}

\author{F.~Bottegoni}
\affiliation{LNESS-Dipartimento di Fisica, Politecnico di Milano, Piazza Leonardo da Vinci 32, 20133 Milano, Italy}

\author{C.~Vergnaud}
\affiliation{Univ. Grenoble Alpes, CEA, CNRS, Grenoble INP (Institute of Engineering Univ. Grenoble Alpes), INAC-Spintec, 38000 Grenoble, France}

\author{T.~Guillet}
\affiliation{Univ. Grenoble Alpes, CEA, CNRS, Grenoble INP (Institute of Engineering Univ. Grenoble Alpes), INAC-Spintec, 38000 Grenoble, France}

\author{A.~Marty}
\affiliation{Univ. Grenoble Alpes, CEA, CNRS, Grenoble INP (Institute of Engineering Univ. Grenoble Alpes), INAC-Spintec, 38000 Grenoble, France}

\author{C.~Beign\'e}
\affiliation{Univ. Grenoble Alpes, CEA, CNRS, Grenoble INP (Institute of Engineering Univ. Grenoble Alpes), INAC-Spintec, 38000 Grenoble, France}

\author{S.~Gambarelli}
\affiliation{Univ. Grenoble Alpes, CEA, INAC-SYMMES, 38000 Grenoble, France}

\author{A.~Picone}
\affiliation{LNESS-Dipartimento di Fisica, Politecnico di Milano, Piazza Leonardo da Vinci 32, 20133 Milano, Italy}

\author{A.~Calloni}
\affiliation{LNESS-Dipartimento di Fisica, Politecnico di Milano, Piazza Leonardo da Vinci 32, 20133 Milano, Italy}

\author{G.~Bussetti}
\affiliation{LNESS-Dipartimento di Fisica, Politecnico di Milano, Piazza Leonardo da Vinci 32, 20133 Milano, Italy}

\author{A.~Brambilla}
\affiliation{LNESS-Dipartimento di Fisica, Politecnico di Milano, Piazza Leonardo da Vinci 32, 20133 Milano, Italy}

\author{L.~Du\`o}
\affiliation{LNESS-Dipartimento di Fisica, Politecnico di Milano, Piazza Leonardo da Vinci 32, 20133 Milano, Italy}

\author{F.~Ciccacci}
\affiliation{LNESS-Dipartimento di Fisica, Politecnico di Milano, Piazza Leonardo da Vinci 32, 20133 Milano, Italy}

\author{P. K.~Das}
\affiliation{CNR-IOM Laboratorio TASC, 34149 Trieste, Italy}

\author{J.~Fujii}
\affiliation{CNR-IOM Laboratorio TASC, 34149 Trieste, Italy}

\author{I.~Vobornik}
\affiliation{CNR-IOM Laboratorio TASC, 34149 Trieste, Italy}

\author{M.~Finazzi}
\affiliation{LNESS-Dipartimento di Fisica, Politecnico di Milano, Piazza Leonardo da Vinci 32, 20133 Milano, Italy}

\author{M.~Jamet}
\email{matthieu.jamet@cea.fr}
\affiliation{Univ. Grenoble Alpes, CEA, CNRS, Grenoble INP (Institute of Engineering Univ. Grenoble Alpes), INAC-Spintec, 38000 Grenoble, France}

\date{\today}% It is always \today, today,

%\begin{abstract}
%Spin-charge interconversion phenomena show a growing interest nowadays in the field of spintronics either to detect spin currents or manipulate the magnetization of ferromagnets. In this respect, spin-orbit coupling is the key ingredient and bismuth appears as a very promising material. However, the spin-charge interconversion efficiency using amorphous or polycrystalline Bi is very low. Here we study the spin-charge interconversion in ultrathin ($0{-}10$ nm) single crystalline Bi films epitaxially grown on Ge(111). Using x-ray diffraction (XRD), scanning tunneling microscopy (STM), and spin- and angle-resolved photoemission spectroscopy (ARPES), we obtain a clear picture of the film morphology, atomic and electronic structures. We then exploit magneto-optical Kerr effect (MOKE), optical spin injection and spin pumping to directly probe the spin-charge interconversion at the Bi/Ge interface. This results to be extremely boosted for $1{-}3$ nm-thick films, thanks to quantum size effects and direct/inverse Rashba-Edelstein effect (REE/IREE). We characterize the IREE efficiency and we estimate an IREE length $\lambda_{\textup{IREE}}\,{=}\,0.35$ nm at the Bi/Ge interface which shows the potential of this interface to manipulate spin currents in Ge. 
%\end{abstract}

%\pacs{72.25.Hg, 72.25.Mk, 85.75.-d, 73.40.Gk, 72.25.Dc}% PACS, the Physics and Astronomy
\maketitle

\textbf{Spin-charge interconversion (SCI) phenomena have attracted a growing interest in the field of spintronics as means to detect spin currents or manipulate the magnetization of ferromagnets. The key ingredients to exploit these assets are a large conversion efficiency, the scalability down to the nanometer scale and the integrability with opto-electronic and spintronic devices. Here we show that, when an ultrathin Bi film is epitaxially grown on top of a Ge(111) substrate, quantum size effects arising in nanometric Bi islands drastically boost the SCI efficiency, even at room temperature. Using x-ray diffraction (XRD), scanning tunneling microscopy (STM) and spin- and angle-resolved photoemission (S-ARPES) we obtain a clear picture of the film morphology, crystallography and electronic structure. We then exploit the Rashba-Edelstein effect (REE) and inverse Rashba-Edelstein effect (IREE) to directly quantify the SCI efficiency using optical and electrical spin injection.}

Bismuth exhibits a series of remarkable electronic properties that have stimulated experimental and theoretical investigations for decades, in particular in electronic transport studies \cite{Fuseya2015,Haas1930,Seebeck1821}. The lattice structure of Bi single layer films resembles that of graphene, while the electronic structure is endowed with a very large spin-orbit coupling of the order of 1.8 eV, which may give origin to topological states \cite{Aguilera2015} or to surface states with a giant Rashba spin-orbit splitting ranging from 0.5 eV \cite{Ohtsubo2012} to 0.8 eV \cite{Koroteev2004}. Bulk Bi has a rhombohedral crystal structure and is a semimetal with a very small indirect band overlap (${\approx}\,38$ meV), resulting in a low charge carrier density compared with conventional metals% (3$\times$10$^{18}$ cm$^{-3}$ at 300 K, 3$\times$10$^{17}$ cm$^{-3}$ at 4 K)
. Electrons exhibit a long Fermi wavelength ($\lambda_{\textup{F}}$) of 40 to 70 nm \cite{Garcia1972,Duggal1969}, which is more than one order of magnitude larger than in typical metals. Moreover, the electron effective mass $m^{*}$ in bismuth amounts to $(0.001{-}0.26)\,m_{\textup{e}}$, depending on the crystalline orientation, with $m_{\textup{e}}$ being the free electron mass \cite{Lin2000a}. The small $m^{*}$ value combined with the long Fermi wavelength facilitates the observation of quantum size effects (QSE) which can drive semimetal to semiconductor (SMSC) transitions in low dimensional systems %. For instance, it was predicted that when the lowest quantized subband of the electron pocket at the $L$ point of the Brillouin zone is raised to an energy higher than the highest hole subband at the $T$ point of the Brillouin zone, a band gap will develop [semimetal-to-semiconductor (SMSC) transition] in nanowires of diameter 20 nm to 45 nm depending on their crystal orientation
\cite{Lin2000b}. %Finally, another important feature of Bi is the existence of surface states with giant Rashba spin-orbit splitting from 0.5 eV \cite{Ohtsubo2012} to 0.8 eV \cite{Koroteev2004}.

To date, the spin properties of Bi films (spin diffusion length $l_{\textup{sf}}$ and spin Hall angle $\theta_{\textup{SH}}$) have shown a large dispersion in experimental values which is probably due to the crystalline state of the material. In amorphous Bi, Emoto \textit{et al.} found: $l_{\textup{sf}}$=8 nm and $\theta_{\textup{SH}}$=0.02 \cite{Emoto2014} whereas, in polycrystalline Bi films, very different values were obtained: $l_{\textup{sf}}\,{=}\,0.11$ nm \cite{Isasa2016}; 2.1 nm \cite{Zhang2015a}; 16 nm \cite{Sangiao2015}; 20 nm \cite{Emoto2016} or 50 nm \cite{Hou2012} at room temperature up to 70 $\mu$m at 2 K \cite{Lee2009} and $\theta_{SHE}\,{=}\,0.00012$ \cite{Emoto2016}; 0.008 \cite{Fan2008}; 0.016 \cite{Sangiao2015}; 0.019 \cite{Hou2012}. Despite the large spin-orbit coupling of Bi, the spin diffusion length is long and the spin-Hall angle $\vartheta_{\textup{SH}}$ quite small, which %e long spin diffusion length
 is detrimental for SCI phenomena or spin-orbit torque switching. %On the other hand, in polycristalline thin Bi films, very different values of $l_{\textup{sf}}$ and $\vartheta_{\textup{SH}}$ have been obtained \cite{Isasa2016,Zhang2015,Sangiao2015,Emoto2016,Hou2012,Lee2009,Fan2008}. %Though very few studies have been conducted on the spin properties of Bi and its potential use in spintronics devices, the long spin diffusion length associated with a very strong spin-orbit coupling make this material a host of choice to discover new spin-dependent phenomena. To date, the spin properties of Bi films (spin diffusion length $l_{sf}$ and spin Hall angle $\theta_{SH}$) have shown a large dispersion in experimental values which is probably due to the crystal quality of the material. In amorphous Bi, Emoto \textit{et al.} found: $l_{sf}$=8 nm and $\theta_{SHE}$=0.02\cite{Emoto2014} whereas, in polycrystalline Bi films, very different values were obtained: $l_{sf}$=0.11 nm\cite{Isasa2016}; 2.1 nm\cite{Zhang2015}; 16 nm\cite{Sangiao2015}; 20 nm\cite{Emoto2016} or 50 nm\cite{Hou2012} at room temperature up to 70 $\mu$m at 2 K\cite{Lee2009} and $\theta_{SHE}$=0.00012\cite{Emoto2016}; 0.008\cite{Fan2008}; 0.016\cite{Sangiao2015}; 0.019\cite{Hou2012}. However, it strongly suggests that the spin diffusion length in single crystalline Bi is very long which is detrimental in using Bi films for SCI or spin-orbit torque switching despite the existence of metallic surface states with giant Rashba splitting. 
For these reasons, much effort has been devoted to the study of Bi-based systems in the ultrathin film limit, where several phenomena were observed such as allotropic transformations \cite{Nagao2004}, the emergence of topologically protected \cite{Ito2016} or superconducting \cite{Weitzel1991} surface states and SMSC transition \cite{Hoffman1993}, which may give rise to a very rich spin physics. 

\begin{figure}[p]
\includegraphics[width=0.7\textwidth]{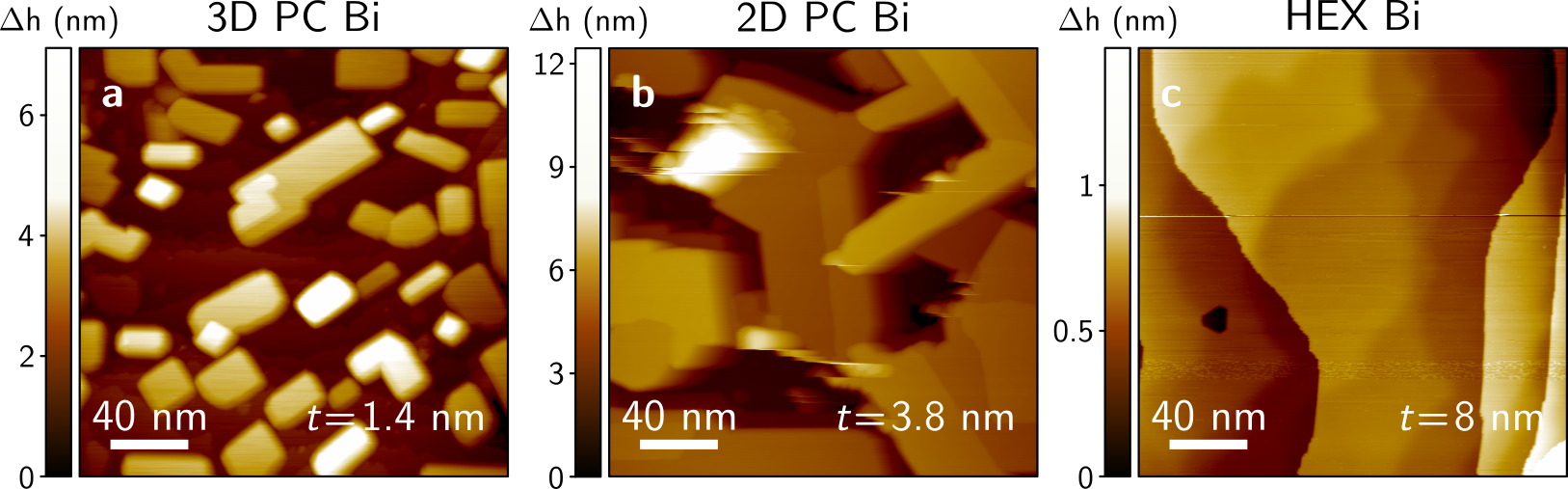} 
\caption{\textbf{Morphology of Bi films grown on Ge(111) as a function of the film thickness $t$.} Typical STM images of \textbf{a}, pseudocubic Bi nanocrystals (3D PC) for $t\,{<}\,3.5$ nm, \textbf{b}, percolated pseudocubic Bi nanocrystals forming a 2D layer (2D PC) for $t\,{=}\,3.8$ nm and \textbf{c}, continuous Bi film with (111)-orientation for $t\,{=}\,8$ nm. The thickness $t$ is calculated starting from the Bi/Ge(111)-$(\sqrt{3}{\times}\sqrt{3}\,)\,R\,30^{\circ}$ wetting layer. %1 ML of Bi was first deposited and annealed on Ge(111) to form the Bi/Ge(111)-$(\sqrt{3}{\times}\sqrt{3}\,)\,R\,30^{\circ}$ wetting layer.
}\label{Fig1}
\end{figure}\

%Bismuth wedges of $t$=0-15 nm for optical measurements and $t$=0-10 nm for spin pumping are grown by molecular beam epitaxy on Ge(111) under ultrahigh vacuum (10$^{-10}$ mbar), at room temperature and a deposition rate of 0.5 \textup{\AA} /s. The wetting layer is the Bi/Ge(111)-($\sqrt{3}\times \sqrt{3})R30^{\circ}$ surface obtained by depositing 1 monolayer of bismuth on Ge(111)-(2$\times$2) annealed at 500$^{\circ}$C for 10 minutes \cite{Hatta2009,Bottegoni2016a}. In parallel, we have conducted spin and angle-resolved photoemission spectroscopy (spin-ARPES) on APE beamline at ELETTRA (Italian synchrotron radiation facility, Trieste) to obtain the spin-dependent electronic structure and scanning tunneling microscopy (STM) to study the film morphology. Spin-ARPES and STM measurements are performed \textit{in-situ} for different bismuth thicknesses ranging from 0 to 10 nm. %The Bi wedges for optical studies are protected by a ZrO$_{2}$(10 nm)/MgO(5 nm) bilayer while, for spin pumping experiments, we deposited a Al(5 nm)/Co(15 nm)/Al(3 nm) trilayer. %IN METHODS
As a first step to investigate SCI in Bi-based low dimensional systems, we analyze the structural and electronic properties of Bi films as a function of the thickness $t$. To this purpose, we have grown ultrathin Bi films ($t\,{\approx}\,0{-}10$ nm) by molecular beam epitaxy (MBE) on a Bi/Ge$(\sqrt{3}{\times}\sqrt{3}\,)\,R\,30^{\circ}$ wetting layer stabilized on a Ge(111) substrate. The Ge substrate exhibits metallic character in the 30-300 K temperature range. In the following, we call the (110)-oriented films the pseudo-cubic (PC) phase  and (111)-oriented films the hexagonal phase (HEX) by analogy with the Bi/Si(111) system \cite{Nagao2004}. Structural characteristics of thin Bi films on Ge(111) were investigated by means of several techniques. The Bi growth proceeds as shown in Fig.~\ref{Fig1} by STM: up to $t\,{=}\,3.5$ nm, we observe the formation of isolated three-dimensional PC flat nanocrystals or nanoplatelets (3D PC phase-Fig.~\ref{Fig1}\textcolor{blue}{$\,$a}); for $3.5\,{<}\,t\,{<}\,4$ nm, the PC nanocrystals start percolating to form a 2D layer (2D PC phase-Fig.~\ref{Fig1}\textcolor{blue}{$\,$b}); for $4\,{<}\,t\,{<}\,5$ nm, there is coexistence of the PC and HEX (PC+HEX) phases and above 5 nm we only observe the single crystalline HEX phase (Fig.~\ref{Fig1}\textcolor{blue}{$\,$c}). This scenario is compatible with that observed in Ref.~\onlinecite{Hatta2009} and is confirmed by \textit{in-situ} Refection High Energy Electron Diffraction (RHEED) and grazing XRD (see Supplementary Information). The latter technique also shows that both PC and HEX films on Ge(111) exhibit the same bulk Bi lattice parameter.

\begin{figure}[p]
\includegraphics[width=0.7\textwidth]{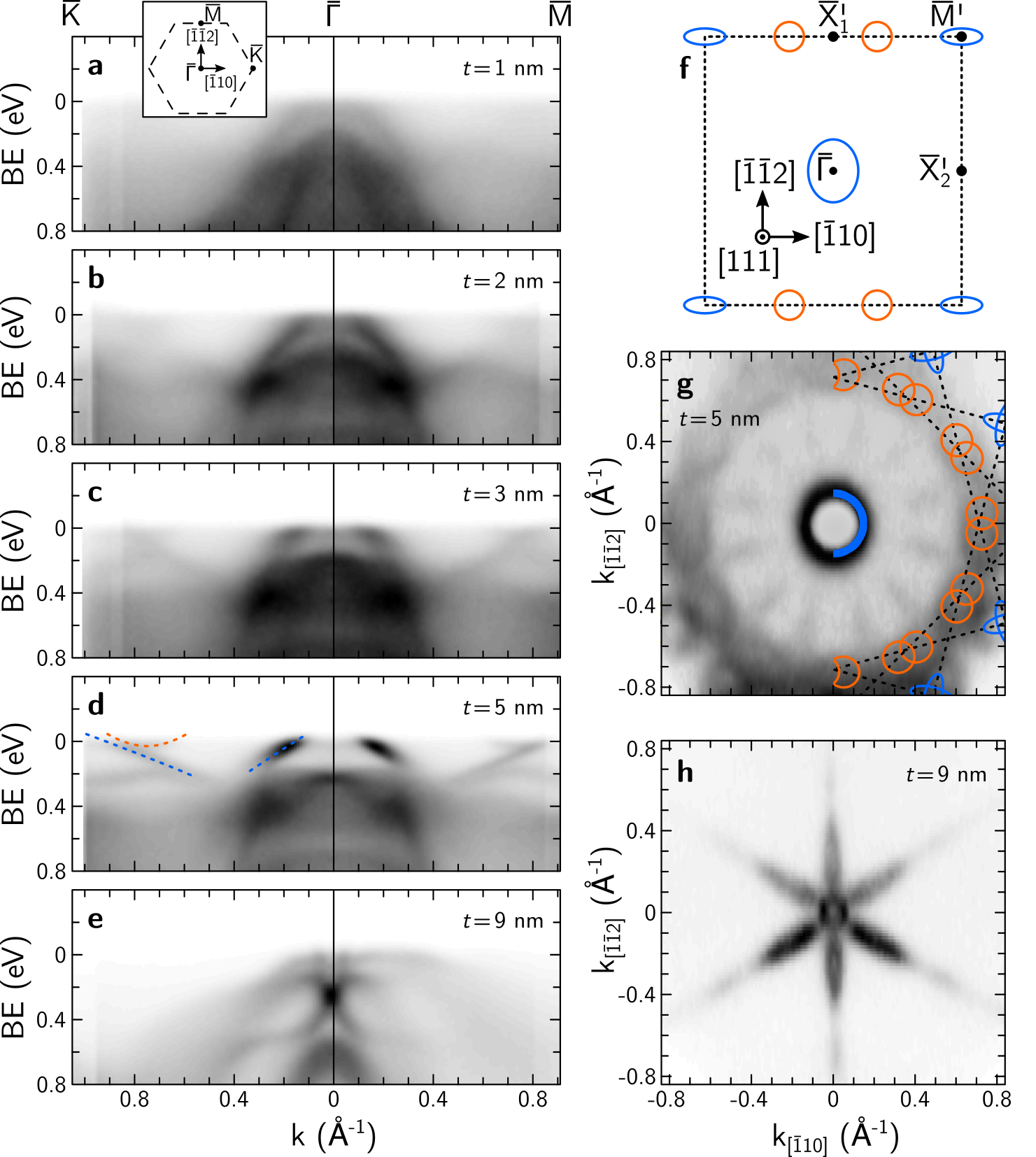} 
\caption{\textbf{Thickness evolution of the band structure of Bi/Ge(111) as obtained by ARPES measurements.} \textbf{a${-}$e}, Band structure along the $\overline{K}{-}\overline{\Gamma}{-}\overline{M}$ directions for $t\,{=}\,1,\,2,\,3,\,5$ and 9 nm of Bi deposited on the Bi/Ge(111)-$(\sqrt{3}{\times}\sqrt{3}\,)\,R\,30^{\circ}$ wetting layer. $\overline{K}$, $\overline{\Gamma}$ and $\overline{M}$ are high symmetry points of the Ge(111) SBZ shown in the inset of panel \textbf{a}. BE and $k$ are the electron binding energy and momentum respectively. Different band structures can be identified depending on the structural phase which are: PC Bi nanocrystals for $t\,{=}\,1{-}3$ nm (3D PC phase), a PC Bi film (2D PC phase) for $t\,{=}\,5$ nm (HEX grains do not give a visible signal) and a (111)-oriented Bi film (HEX phase) for $t\,{=}\,9$ nm. \textbf{f}, Schematics of the Bi(110) 2D Fermi surface according to Ref.~\onlinecite{Agergaard2001}. $\overline{M}'$, $\overline{\Gamma}'\,{=}\,\overline{\Gamma}$ and $\overline{X_{1}}'$ and $\overline{X_{2}}'$ are the high symmetry points of the Bi(110) SBZ. Blue lines (around $\overline{\Gamma}$ and $\overline{M}'$) correspond to hole states while orange ones (near $\overline{X_{1}}'$) to electron states. They are also reported in \textbf{d}. $k_{[\bar{1}10]}$, $k_{[\bar{1}\bar{1}2]}$ and $k_{[111]}$ are the basis vectors of the reciprocal space. % used to calculate the confinement effect (see text). 
\textbf{g}, Experimental 2D Fermi surface for $t\,{=}\,5$ nm reproduced (in orange and blue lines) by superimposing 6 times the Fermi surface of bulk Bi(110) in \textbf{f}. \textbf{h}, 2D Fermi surface of single crystalline Bi(111) for $t\,{=}\,9$ nm.}\label{Fig2}
\end{figure}

The electronic properties of \textit{in situ} grown Bi/Ge(111) samples have been characterized by means of ARPES with spin resolution (S-ARPES) at the APE beamline of the ELETTRA synchrotron radiation facility. In Figs.~\ref{Fig2}\textcolor{blue}{$\,$a${-}$e}, the ARPES spectra collected along the $\overline{K}{-}\overline{\Gamma}{-}\overline{M}$ directions of the Ge(111) surface Brillouin zone (SBZ; sketched in the inset of Fig.~\ref{Fig2}\textcolor{blue}{$\,$a}) are reported as a function of the Bi thickness. At the early stages of growth, in the 3D PC regime ($t\,{=}\,1{-}3.5$ nm), we observe states crossing the Fermi level ($E_{\textup{F}}$) around the $\overline{\Gamma}$ point with a hole character. In analogy with what Bian et al. observed on thin Bi/Si(111) \cite{Bian2009}, we conclude that these are surface states with a very short spatial extension of only 2 Bi bilayers ($1~\textup{BL}\,{=}\,3.28~\textup{\AA}$), as shown in Fig.~4 of Ref.~\onlinecite{Bian2009}. In this regime, we do not clearly observe other surface or bulk states close to $E_{\textup{F}}$. 
For $t\,{=}\,5$ nm, the band structure has evolved and clearly shows occupied states around $k_{\parallel}\,{=}\,0.7$ $\textup{\AA}^{-1}$ along both the $\overline{\Gamma}{-}\overline{K}$ and $\overline{\Gamma}{-}\overline{M}$ directions. %Finally, for $t\,{=}\,9$ nm, we find the band structure of single crystalline Bi(111)\cite{Hirahara2006}. IT IS WRITTEN BELOW
At this thickness, it is worth comparing our experimental data (Fig.~\ref{Fig2}\textcolor{blue}{$\,$g}) with the calculated 2D Fermi surface of Bi(110) (Fig.~\ref{Fig2}\textcolor{blue}{$\,$f}) \cite{Agergaard2001,Koroteev2004}. In the former, we observe a dark ring around $\overline{\Gamma}$ and 12 elongated low-intensity rings centered around $k_{\parallel}\,{=}\,0.7$ $\textup{\AA}^{-1}$. This Fermi surface can be reproduced by considering the ARPES results of Agergaard et al. on the (110) surface of bulk Bi \cite{Agergaard2001} and the first principles calculations of Koroteev et al. \cite{Koroteev2004}. Indeed, six equivalent growth orientations of the 2-fold symmetric Bi(110) surface are detected on the 6-fold symmetric Ge(111) surface.  %This explains the similarity in the measured band structures along the $\overline{\Gamma}{-}\overline{K}$ and the $\overline{\Gamma}{-}\overline{M}$ directions (Figs.~\ref{Fig2}\textcolor{blue}{$\,$a${-}$d}). 
The elongated rings correspond to the surface electronic states along the $\overline{M}'{-}\overline{X_{1}}'$ direction of the Bi(110) SBZ. They are also reported in Fig.~\ref{Fig2}\textcolor{blue}{$\,$d} along with the surface hole states at $\overline{M}'$ which cross the Fermi level at $k_{\parallel}\,{=}\,0.9$ $\textup{\AA}^{-1}$. 
Finally, for $t\,{=}\,9$ nm (Figs.~\ref{Fig2}\textcolor{blue}{$\,$e,$\,$h}), the band structure is the one of single crystalline bulk Bi(111) \cite{Hirahara2006}. We note here that HEX Bi films grow in registry with Ge(111) (for further details about the crystallography of thin Bi/Ge(111) films, see the Supplementary Information). 

\begin{figure}[p]
\includegraphics[width=0.7\textwidth]{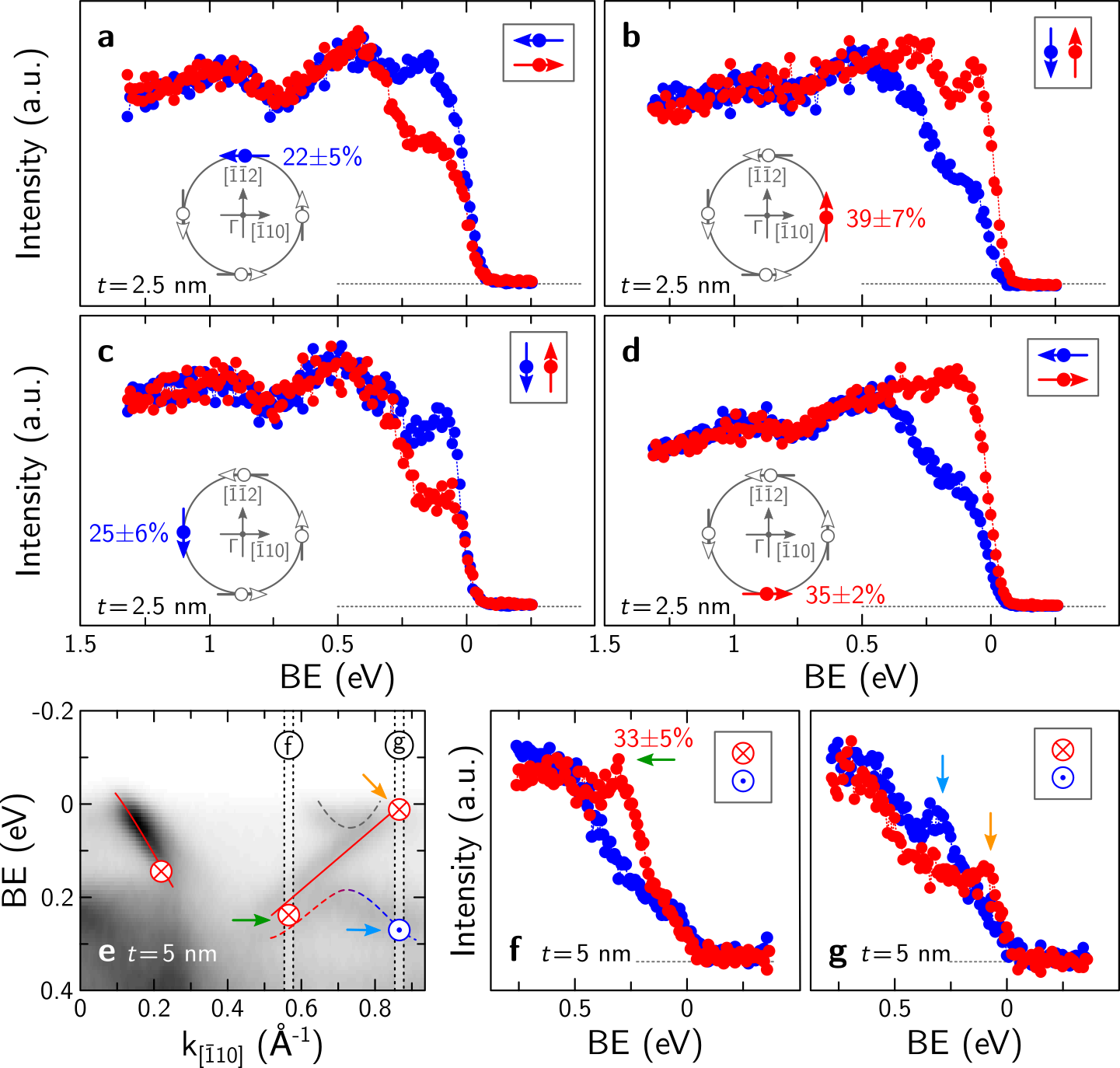} 
\caption{\textbf{Spin-texture of metallic surface states.} \textbf{a${-}$d}, Spin polarization recorded along the $\overline{\Gamma}$ states showing the helical spin texture, for $t\,{=}\,2.5$ nm. The numbers are the net spin polarization values. %and signs are given with respect to $k_{x}$ and $k_{y}$ directions.
\textbf{e}, Band structure along $\overline{\Gamma}{-}\overline{K}$ for $t\,{=}\,5$ nm. Red solid lines are $\overline{\Gamma}$ ($k_{[\bar{1}10]}\,{\approx}\,0.1$ $\textup{\AA}^{-1}$) and $\overline{M}'$ ($k_{[\bar{1}10]}\,{\approx}\,0.7$ $\textup{\AA}^{-1}$) hole states of Bi(110) respectively. Dotted lines crossing the Fermi level correspond to the electron pockets near the $\overline{X_{1}}'$ point shown in Figs.~\ref{Fig2}\textcolor{blue}{$\,$f,$\,$g}. \textbf{f,$\,$g}, Spin polarization of $\overline{M}'$ states.}\label{Fig3}
\end{figure}

The results of S-ARPES are shown in Fig.~\ref{Fig3}. In Figs.~\ref{Fig3}\textcolor{blue}{$\,$a${-}$d}, for $t\,{=}\,2.5$ nm, we probe the surface states around $\overline{\Gamma}$: they exhibit a counterclockwise helical spin texture with a spin polarization $P$ up to 40\%. The spin-momentum locking is due to the strong Rashba spin-orbit coupling in the surface states. In Figs.~\ref{Fig3}\textcolor{blue}{$\,$e${-}$g}, we show that the states at $\overline{M}'$ for $t\,{=}\,5$ nm are also spin polarized ($P\,{\approx}\,30\%$), but they have a clockwise spin helicity, as predicted by Pascual et al. \cite{Pascual2004}. The conclusion of this analysis is that both $\overline{\Gamma}$ and $\overline{M}'$ states can participate to SCI, but they would provide opposite contributions, since they are both holes states and have opposite spin helicities.

\begin{figure}[p]
\includegraphics[width=0.7\textwidth]{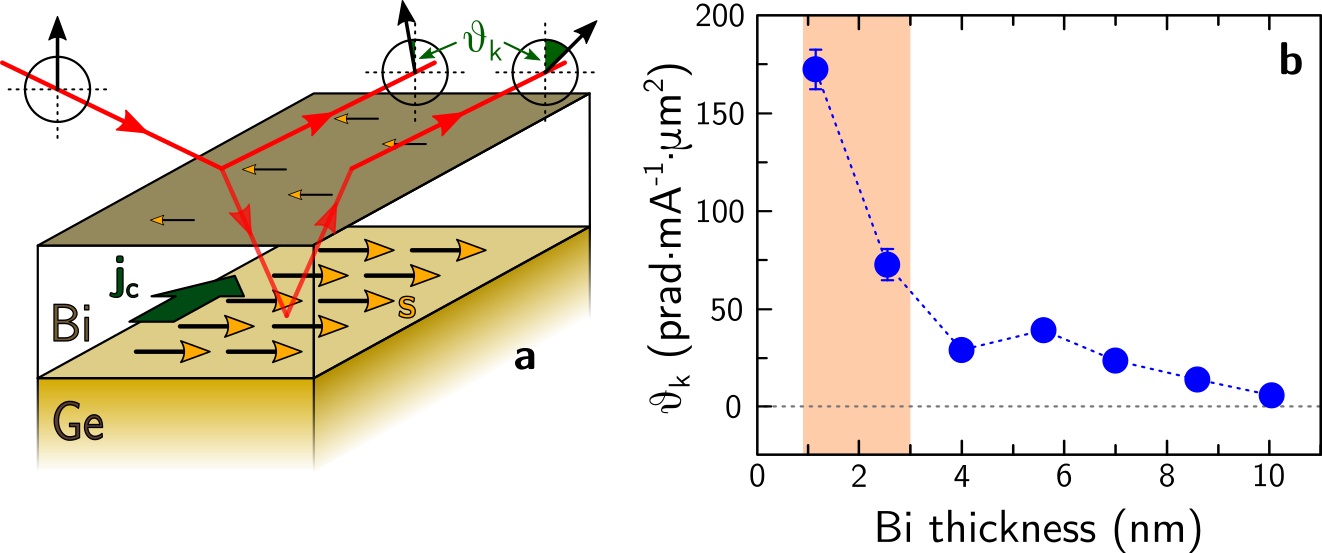} 
\caption{\textbf{Charge-to-spin conversion in Bi films probed by MOKE.} \textbf{a}, Schematics of the experimental setup for charge-to-spin conversion. An electrical current flows into or at the surface of the Bi layer and is converted into a spin accumulation at the top and bottom surfaces. The spin accumulation at the top surface is detected by longitudinal Kerr effect. \textbf{b}, Kerr angle $\vartheta_{\textup{k}}$ detected as a function of the Bi thickness.}\label{Fig4}
\end{figure}

Based on the accurate knowledge of the atomic and electronic structures of ultrathin Bi films, we then perfomed SCI measurements as a function of the Bi thickness. Charge-to-spin conversion phenomena can be directly probed by exploiting MOKE. In this respect, the detection of an electrically-induced spin accumulation in metals is particularly challenging and it has been limited to low temperature ranges \cite{Stamm2017}. In our case, we exploit longitudinal MOKE (see Fig.~\ref{Fig4}\textcolor{blue}{$\,$a}): an electrical current flows in 1.5 mm-wide Bi/Ge(111) stripes of constant Bi thickness and we detect the Kerr rotation signal coming from the Bi film with a double modulation technique at room temperature (see Supplementary Information). %For $0\,{<}\,t\,{<}\,3$ nm, since the nanocrystals are semiconducting, most of the current flows at the Bi/Ge interface, where the conductivity of the Rashba electron gas is large \cite{Sinova2015}. As a consequence, the REE generates an in plane spin accumulation, with a spin polarization perpendicular to the current density vector. In this thickness range we detect a very large Kerr signal, which is proportional to the electrically-induced spin density at the Bi/Ge interface (see Fig.~\ref{Fig4}$\,$b), since the absorption length $\alpha$ of the incident light ($\alpha\,{=}\,16$ nm for $\lambda\,{=}\,671$ nm) is much larger than the nanocrystal height \cite{Hagemann1975}. On the contrary, for $t\,{>}\,3$ nm, nanocrystals become gradually conducting and the electrical current flows in both the island interfaces, causing opposite spin accumulations, which tend to cancel each other. Hence, the Kerr signal drastically decreases and vanishes in correspondence at the PC-hexagonal Bi transition. On the other hand, it is well known that the spin-Hall angle in Bi is quite small \cite{Emoto2014,Sangiao2015,Emoto2016,Fan2008,Hou2012}, so that the spin-Hall effect could generate only a negligible electrically-induced spin accumulation compared to the one generated by the REE.
As shown in Fig.~\ref{Fig4}\textcolor{blue}{$\,$b}, up to $t\,{=}\,3$ nm we detect a large Kerr signal $\vartheta_{\textup{k}}$, which results from the electrically-induced spin accumulation in Bi, whereas $\vartheta_{\textup{k}}$ rapidly decreases as the Bi thickness is increased. 

\begin{figure}[p]
\includegraphics[width=0.7\textwidth]{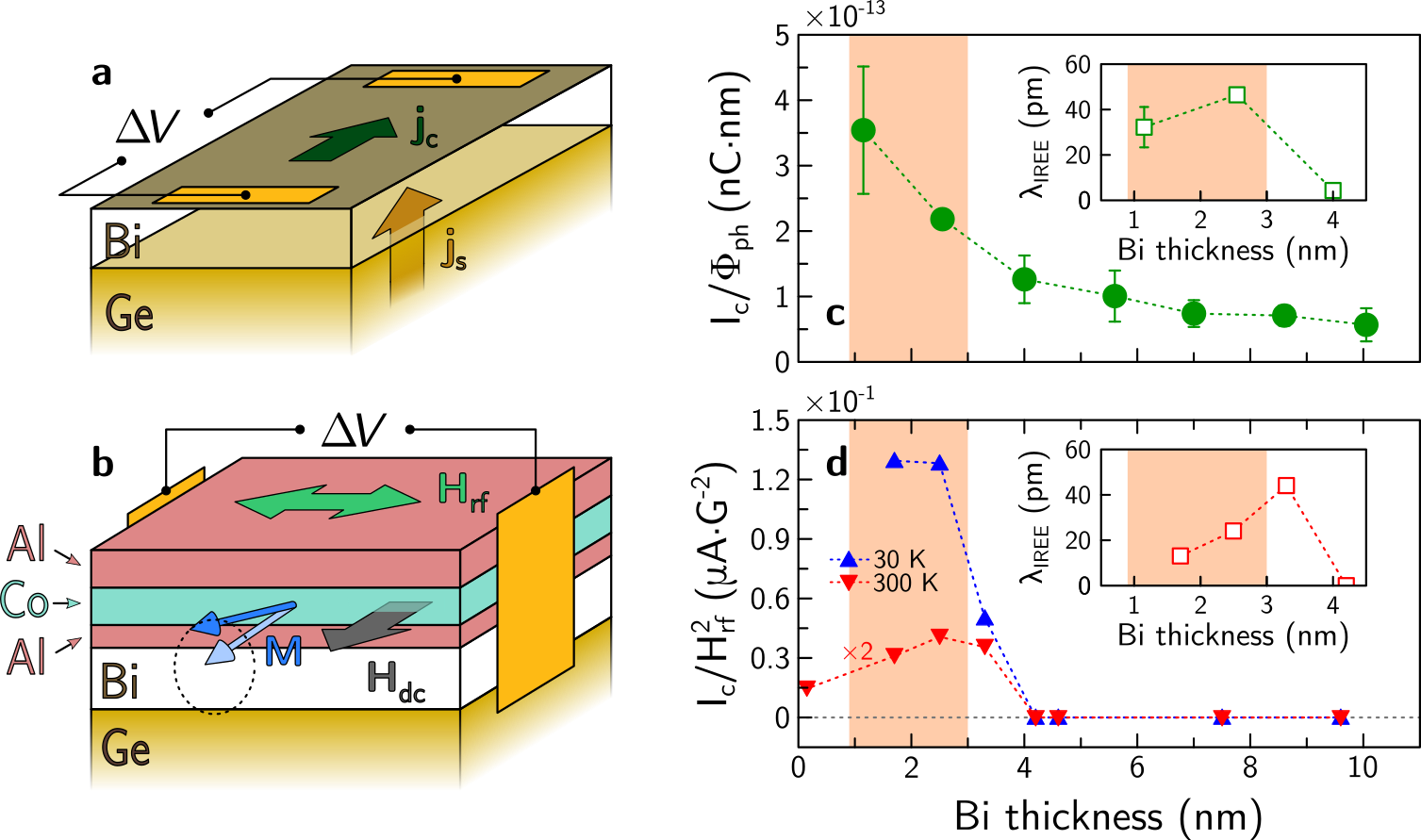} 
\caption{\textbf{Spin-to-charge conversion efficiency probed by either optical or electrical spin injection.} \textbf{a, b}, Schematic drawings showing the experimental geometries used for optical spin orientation and spin pumping measurements respectively (for more details refer to the Methods and Supplementary Information). \textbf{c}, Bi-thickness dependence of the spin-to-charge conversion efficiency $I_{\textup{C}}/\Phi_{\textup{ph}}$ at room temperature using optical spin orientation. $I_{\textup{C}}=\Delta V/R$ is the generated charge current where $\Delta V$ is the voltage measured in open circuit conditions upon illumination with circularly polarized light and $R$ is the electrical resistance between the two contacts estimated using four-probe resistance geometry. The charge current is normalized to the excitation signal, i.e., the photon flux $\Phi_{\textup{ph}}$. Inset: $\lambda_{\textup{IREE}}$ values deduced at room temperature. \textbf{d}, Bi-thickness dependence of the spin-to-charge conversion efficiency $I_{\textup{C}}/H_{\textup{rf}}^{2}$ at 30 K and room temperature using spin pumping. $I_{\textup{C}}=\Delta V/R$ is the generated charge current where $\Delta V$ is the voltage measured in open circuit conditions at the ferromagnetic resonance of the Co electrode and $R$ is the electrical resistance between the two contacts. The charge current is normalized to the excitation signal, i.e., the radiofrequency power proportional to $H_{\textup{rf}}^{2}$. Inset: $\lambda_{\textup{IREE}}$ values deduced at room temperature.}\label{Fig5}
\end{figure}

The same qualitative behavior is found when the spin-to-charge conversion generated by a spin current is investigated by either optical spin orientation in Ge \cite{Bottegoni2014} or spin pumping from a ferromagnet. Optical spin orientation allows obtaining a spin accumulation with in-plane polarization in Ge by shining circularly polarized light on the sample at a grazing incidence \cite{Pierce1976}. The spin-polarized electrons then diffuse into the Bi film (Fig.~\ref{Fig5}\textcolor{blue}{$\,$a}) \cite{Isella2015}. As an alternative to optical spin orientation, we can also inject a spin current by spin pumping from an Al(5 nm)/Co(15 nm)/Al(3 nm) stack grown on top of Bi at the ferromagnetic resonance of the Co layer (Fig.~\ref{Fig5}\textcolor{blue}{$\,$b}) \cite{Ando2011}. In both cases the spin current generates a transverse charge current, which is detected as a voltage $\Delta V$ measured between two electrodes deposited across the Bi film in open circuit conditions. Further details on both techniques can be found in the Methods section and in the Supplementary Information.
In Fig.~\ref{Fig5}\textcolor{blue}{$\,$c}, we show the results for optical spin orientation measurements: the signal is larger in the 3D PC regime, whereas for $t\,{>}\,3$ nm, it decreases. Similarly, with spin pumping (Fig.~\ref{Fig5}\textcolor{blue}{$\,$d}) at 30 K and 300 K we observe a sharp decrease of the signal above $t\,{=}\,3$ nm, with no detectable signal for $t\,{>}\,4$ nm. The signal at 30 K is roughly one order of magnitude larger compared to that at room temperature. %Moreover, for $t\,{=}\,50$ nm, we still detect no signal which is indicative of negligible bulk spin-to-charge conversion by ISHE and a long spin diffusion length $l_{\textup{sf}}\,{>}\,50$ nm. 
To summarize, in all the SCI experiments we observe a conversion signal for $t\,{<}\,4$ nm, when Bi nanocrystals are present at the surface of Ge(111), whereas the signal is drastically attenuated for the other morphologies.

We can exclude that the Bi/Ge(111)-$(\sqrt{3}{\times}\sqrt{3}\,)\,R\,30^{\circ}$ wetting layer significantly contributes to SCI since Aruga et al. demonstrated both experimentally and theoretically the absence of spin-polarized states at the Fermi level \cite{Aruga2015}. Hence, the interconversion takes place within or at the surface of nanocrystals of a given height $h$ and lateral size $a\,{=}\,\sqrt{S}$, being $S$ the nanocrystal area. STM images show that the lateral size $a$ of nanocrystals is comparable to the Fermi wavelength, and since $h\,{\ll}\,a$, quantum confinement effects play a fundamental role in determining the spin-transport properties of the system. Indeed, due to the low effective mass of electrons in bulk Bi, the spacing between discrete energy levels can be large enough to open a bandgap $E'_{\textup{g}}$ in Bi nanocrystals \cite{Zhang2000}. %This SMSC transition can be shown by using the following expression of the effective bandgap $E'_{g}$ between the lowest $L$-point electron subband and the highest $T$-point hole subband\cite{Zhang2000}:
The calculation of $E'_{\textup{g}}$ in the Supplementary Informations gives a SMSC transition (i.e., $E'_{\textup{g}}\,{\geq}\,0$) for $a\,{\leq}\,50$ nm. Hence, nanocrystals of lateral size $a\,{\leq}\,50$ nm and thicker than 4 BL (meaning that opposite surface states do not overlap each other \cite{Bian2009}) are expected to be semiconducting. %Notably, the nanocrystals observed by STM satisfies these conditions when $t\,{<}\,3$ nm (further details in Supplementary Informations).
%
%The SMSC transition and the long spin diffusion length $l_{\textup{sf}}$ in Bi allows us to interpret the SCI results. 
From an extensive analysis of STM images (see the Supplementary Information), for $t\,{<}\,0.9$ nm, the majority of PC nanocrystals are less than 4 BL-thick, the top and bottom surface states overlap and the corresponding SCIs, being of opposite sign, cancel each other. For $0.9\,{<}\,t\,{<}\,3$ nm, the nanocrystals satisfy the conditions for a SMSC transition. The  surface states do not overlap and are electrically separated because the presence of a bandgap increases the bulk resistance. As a consequence, SCI at each interface is allowed and no communication channel between opposite interfaces is present. We can observe a net SCI signal. This thickness range is reported as a shaded area in Figs.~\ref{Fig4}\textcolor{blue}{$\,$b} and ~\ref{Fig5}\textcolor{blue}{$\,$c,$\,$d}, and it nicely fits the thickness range where we experimentally observe a conversion signal. This interpretation is in agreement with the increase of the spin-pumping signal at low temperature. For $0.9\,{<}\,t\,{<}\,3$ nm, Bi nanocrystals exhibit a bandgap and the number of thermally excited electrons across this bandgap is lower at 30 K than at room temperature, which increases the bulk resistance. Spin-to-charge conversion takes place at the $\overline{\Gamma}$ states in Figs.~\ref{Fig2}\textcolor{blue}{$\,$a${-}$c} due to their helical spin texture shown in Figs.~\ref{Fig3}\textcolor{blue}{$\,$a${-}$d}, thus it can be attributed to the IREE. Concerning Kerr measurements, in this thickness range most of the current flows at the Bi/Ge interface, where the conductivity of the Rashba electron gas is large \cite{Sinova2015}. As a consequence, the REE generates an in-plane spin accumulation, with a spin polarization perpendicular to the current density vector. We detect a very large Kerr signal (Fig.~\ref{Fig4}\textcolor{blue}{$\,$b}), which is proportional to the electrically-induced spin density at the Bi/Ge interface, since the absorption length $\alpha$ of the incident light ($\alpha\,{=}\,16$ nm for $\lambda\,{=}\,691$ nm) is much larger than the nanocrystal height \cite{Hagemann1975}. Finally, for $t\,{>}\,3$ nm, nanocrystals exhibit lateral sizes larger than $\lambda_{\textup{F}}$ and start percolating. This reduces and finally suppresses quantum confinement at room temperature. In these conditions, spin-polarized electrons diffuse in the entire film thickness, and being $h\,{<}\,l_{\textup{sf}}$, the spin-to-charge conversions at both interfaces compensate each other reducing the signal down to zero. Similarly, for charge-to-spin conversion, when nanocrystals become gradually conducting, electrical currents flow at both interfaces, causing opposite spin accumulations, which tend to cancel each other. Hence, the Kerr signal drastically decreases. 
In a simple model, the effect of quantum confinement on SCI experiments is equivalent to the effect of a variable bulk resistance $R_{\textup{B}}$ electrically connecting the top and bottom metallic surface states of resistance $R_{\textup{S}}$. Following Ref.~\onlinecite{Zhang2000}, for $t<$3 nm, quantum confinement leads to $R_{\textup{B}}\gg R_{\textup{S}}$ and surface states are electrically insulated from each other. We can observe SCI signals. On the other hand, for $t>$3 nm, $R_{\textup{B}}\approx R_{\textup{S}}$ and the charge currents in the top and bottom surface states are shunted through the bulk reducing and cancelling SCI signals. We could not detect any spin-to-charge conversion by spin pumping for $t\,{=}\,50$ nm. This is indicative of negligible bulk spin-to-charge conversion by inverse spin Hall effect and a long spin diffusion length $l_{\textup{sf}}\,{>}\,50$ nm.
Starting from $t\,{=}\,3$ nm, $\overline{M}'$ surface states at $E_{\textup{F}}$ develop at the surface of Bi nanocrystals and films as shown in Fig.~\ref{Fig2}. They exhibit a hole character and a spin chirality opposite to the one of $\overline{\Gamma}$ states (Fig.~\ref{Fig3}), thus also contributing to the decrease of conversion signals.

Spin-to-charge measurements allow extracting the figure of merit of the conversion occurring at the interfaces (insets of Fig.~\ref{Fig5}\textcolor{blue}{$\,$c,d}). It corresponds to the IREE length $\lambda_{\textup{IREE}}\,{=}\,j_{\textup{c}}^{2\textup{D}}/j_{\textup{s}}$, where $j_{\textup{c}}^{2\textup{D}}$ is the 2D charge current density (in $\textup{A\,m}^{-1}$) generated by the 3D spin current density $j_{\textup{s}}$ (in $\textup{A\,m}^{-2}$) \cite{Rojas-Sanchez2013}. The methods to calculate $\lambda_{\textup{IREE}}$ from optical spin orientation and spin pumping measurements are detailed in the Supplementary Information: there, we assume that the conversion occurs only in PC nanocrystals that fulfill the conditions $h\,{>}\,4$ BL (surface states do not overlap) and $a\,{\leq}\,50$ nm (nanocrystals are semiconducting with high bulk resistance). The fraction of the sample surface corresponding to these nanocrystals is given by the analysis of STM images. In the conversion process, the transverse charge current generated at the nanocrystal interfaces is transfered to the conducting Ge substrate for optical spin orientation experiments and to the Al/Co/Al metallic trilayer for spin pumping experiments. It is then detected as a voltage in open circuit conditions. We obtain a maximum calculated value of ${\approx}\,50$ pm by both optical spin orientation and spin pumping for $t\,{\approx}\,3$ nm. It shows that the spin-to-charge conversion efficiencies are comparable at the Bi/Ge and Bi/Al interfaces and that SCI occurs into the $\overline{\Gamma}$ surface states of Bi regardless the material at the interface. It is important to note that such a calculation is performed under the assumption that the bulk resistance of the nanocrystals is large enough to avoid spin diffusion between the two interfaces. If a lower resistance value were taken into account, the $\lambda_{\textup{IREE}}$ value would be drastically larger, so that 50 pm represents a lower bound estimation of the spin-to-charge interconversion efficiency. This $\lambda_{\textup{IREE}}$ value is comparable to the ones obtained at different Rashba interfaces such as Ag/Bi (100-300 pm) \cite{Rojas-Sanchez2013,Zhang2015b}, Ag/Sb (30 pm) \cite{Zhang2015b} or Cu/Bi (9 pm) \cite{Isasa2016}. For Rashba interfaces, $\lambda_{\textup{IREE}}\,{=}\,\alpha_{\textup{R}}\tau/\hbar$ where $\alpha_{\textup{R}}$ is the Rashba coefficient and $\tau$ is the momentum relaxation time in the interface states. From Ref.~\onlinecite{Koroteev2004}, we can estimate $\alpha_{\textup{R}}\,{\approx}\,1.5\,{\times}\,10^{-10}$ eV$\,$m at the Bi(110) surface assuming nearly free electrons in surface states, which gives $\tau\,{\approx}\,0.2$ fs. This value is of the same order of magnitude as $\tau$ values obtained at other Rashba interfaces.\cite{Rojas-Sanchez2013,Shen2014,Isasa2016}. 

To summarize, we carried out careful structural and electronic characterizations of Bi thin films epitaxially grown on Ge(111). SCI in Bi layers was investigated by Kerr effect, optical spin orientation and spin pumping. In all three techniques, a conversion signal was only observed in the $t\,{=}\,1{-}3$ nm thickness range corresponding to the presence of Bi(110) nanocrystals. We thus interpreted the results as a consequence of QSE and SCI at the surface of semiconducting nanocrystals by (I)REE. Eventually, we found a $\lambda_{\textup{IREE}}$ value as high as 50 pm at the Bi/Ge interface which shows the potential of this interface to manipulate spin currents in Ge. Since this is also the first evidence of SCI triggered by QSE, our results paves the way for the exploitation of QSE to tune SCI, and open a new route to manipulate spin currents in Ge by Rashba effect at the interface with a metal \cite{Oyarzun2016}.\\

\noindent \textbf{Acknowledgments}\\ \noindent
{\footnotesize This work was supported by the ANR-16-CE24-0017 project TOP RISE and by the Laboratory of Excellence LANEF of Grenoble (ANR-10-LABX-51-01). The authors would also like to acknowledge Dr. Henri Jaffr\`es for fruitful discussions.\\}

\noindent \textbf{Author contributions}\\ \noindent
{\footnotesize M.J., F.B., L.D., F.C. and M.F. coordinated the entire project. C.Z. and M.T.D. equally participated to the work. A.P. and A.B. performed STM measurements. M.T.D., S.G., C.V., T.G., A.M., C.B. and M.J. performed XRD and  spin pumping measurements, and provided the samples for optical and electrical spin injection. C.Z., F.B., C.V., A.P., A.C., G.B., A.B., M.F., P.K.D., J.F., I.V. and M.J. carried out ARPES and S-ARPES measurements. C.Z. and F.B. carried out MOKE and optical spin injection measurements. C.Z., M.T.D, F.B., A.M., A.P., A.C. and M.J. performed the data analysis. All the authors contributed to the writing of the manuscript.}\\

%\noindent \textbf{Additional information}\\ \noindent
%{\footnotesize\textbf{Supplementary Information} accompanies this paper.\\ \noindent
%\textbf{Correspondence} should be addressed to M.J.\\}

\noindent \textbf{Methods}\\ \noindent
{\footnotesize\textbf{Sample preparation.}
ARPES and STM measurements were performed \textit{in-situ} for bismuth thicknesses ranging from 0 to 10 nm. Bismuth was grown by molecular beam epitaxy on Ge(111) under ultrahigh vacuum (10$^{-10}$ mbar), at room temperature and a deposition rate of 0.5 \textup{\AA}/s. The wetting layer was the Bi/Ge(111)-$(\sqrt{3}{\times}\sqrt{3}\,)\,R\,30^{\circ}$ surface obtained by depositing 1 ML of bismuth on Ge(111)-$(2{\times}2)$ annealed at $500^{\circ}$C for 10 minutes \cite{Hatta2009,Bottegoni2016a}. $0{-}10$ nm Bi wedges for optical and electrical measurements were grown in the same way. The Bi wedges for optical studies were protected by a ZrO$_{2}$(10 nm)/MgO(5 nm) bilayer grown \textit{in-situ}. The first nanometer of MgO was deposited using e-beam evaporation at a very low rate (0.025 \textup{\AA}/s) in order to limit the oxygen pressure in the MBE chamber and avoid partial oxidation of the Bi film. The last 4 nanometers were deposited at a rate of 0.25 \textup{\AA}/s. The ZrO$_{2}$ layer is grown \textit{in-situ} by RF sputtering. For spin pumping experiments, we deposited \textit{in-situ} an Al(5 nm)/Co(15 nm)/Al(3 nm) trilayer.\\

\noindent\textbf{(S-)ARPES measurements.}
(S-)ARPES measurements were performed using \textit{p}-polarized synchrotron radiation at the APE beamline of Elettra with a photon energy $h\nu\,{=}\,50$ eV. The hemispherical electron energy and momentum analyzer (Scienta DA30) is equipped with two very low-energy electron diffraction (V-LEED)-based spin polarimeters. We probe the in-plane components of the spin polarization, and the spin-detection efficiency was corrected using a Sherman function ($S\,{=}\,0.3$), determined by comparison with the known spin polarization of the Rashba-split surface states measured on the Au(111) surface. The spin polarization $P$ is extracted as: $P\,{=}\,[I^{+}\,{-}\,I^{-}]/[S\,{\times}\,(I^{+}\,{+}\,I^{-})]$, where $I^{+(-)}$ is the V-LEED scattering intensity measured for the V-LEED target magnetization in the positive (negative) direction. The detailed description of the S-ARPES setup can be found in Ref.~\onlinecite{Bigi2017}.\\

\noindent\textbf{MOKE measurements.}
Electrically-induced spin accumulation in Bi is detected by means of Longitudinal MOKE (L-MOKE). We used a 691 nm-continuous wave laser as a light source. The \textit{s}-polarized light was focused on the sample surface with an average polar angle $\vartheta\,{=}\,45^{\circ}$, a spotsize of diameter 5 $\mu$m and an optical power of $W\,{\approx}\,125$ $\mu$W. The reflected light beam passed through a photo-elastic modulator, which modulated the circular polarization of the light at 50 kHz, and a polarizer, before being  collected by a photodiode as described in Ref.~\onlinecite{Yang1993}. We recorded the second harmonic of the signal with a first lock-in amplifier and normalized the result to the sample reflectivity to obtain the pure ellipticity signal. In order to further increase the signal-to-noise ratio, we modulated the charge current at 0.3 Hz and extracted the optical signal with a second lock-in amplifier. Further details are given in the Supplementary Information.\\

\noindent\textbf{Optical spin orientation.}
A circularly polarized laser beam ($\lambda\,{=}\,740$ nm) was focused on the sample with a spotsize diameter $s\,{\approx}1.5$ $\mu$m and an optical power $W\,{\approx}\,19$ mW. Optical spin injection generates a spin-oriented population of electrons in the Ge conduction band, with a spin polarization $P\,{\approx}\,8\%$ \cite{Rioux2010} parallel to the light wavevector inside Ge \cite{Pierce1976}. Since the experimental geometry is sensitive to the in-plane component of the spin polarization \cite{Bottegoni2017}, we had the laser beam partially filling off-axis a 0.65 numerical aperture objective, focusing the light on the sample with a polar angle $\vartheta\,{\approx}\,20^{\circ}$. The resulting electromotive force is measured under open-circuit conditions. The circular polarization is modulated by a photoelastic modulator at 50 kHz and the signal is demodulated by a lock-in amplifier. Furthermore, to increase the signal-to-noise ratio, we also modulate the light intensity at 21 Hz with a chopper and the signal is extracted by a second lock-in amplifier.\\

\noindent\textbf{Spin pumping.}
 A transverse radiofrequency field $\bf{H}_{\textup{rf}}$, generated at the center of a cylindrical X-band resonator cavity ($f\,{=}\,9.7$ GHz, $\textup{TE}_{011}$ mode), triggers the ferromagnetic resonance of the Co layer and spin pumping. The charge current is given by $I_{\textup{C}}=\Delta V/R$, where $R$ is the resistance measured between the two voltage probes. To remove the Seebeck contribution to the signal at room temperature, we consider: $I_{\textup{C}}\,{=}\,(I_{\textup{C}}^{+\bf{H_{\textup{dc}}}}\,{-}\,I_{\textup{C}}^{\bf{-H_{\textup{dc}}}})/2$ where $\bf{H}_{\textup{dc}}$ is the DC magnetic field applied in the film plane \cite{Shiomi2014}.}\\

\end{document}